\documentclass{aipproc}
\layoutstyle{6x9}

\newcommand{\pt}{{\partial}}

\newcommand{\be}{\begin{eqnarray}}
\newcommand{\ee}{\end{eqnarray}}

 \newcommand{\beq}{\begin{equation}}
\newcommand{\eeq}{\end{equation}}
\newcommand{\bea}{\begin{eqnarray}}
\newcommand{\eea}{\end{eqnarray}}
\newcommand{\beqn}{\begin{eqnarray}}
\newcommand{\eeqn}{\end{eqnarray}}
\newcommand{\beas}{\begin{eqnarray*}}
\newcommand{\eeas}{\end{eqnarray*}}

\newcommand{\bquo}{\begin{quote}}
\newcommand{\enqu}{\end{quote}}

\newcommand{\tN}{\tilde N}

\newcommand{\gsim}{\lower.7ex\hbox{$
\;\stackrel{\textstyle>}{\sim}\;$}}
\newcommand{\lsim}{\lower.7ex\hbox{$
\;\stackrel{\textstyle<}{\sim}\;$}}

\def\d{\partial}

\def\2{{1\over 2}}
\def\ntwo{${\mathcal N}=2\;$}

\def\none{${\mathcal N}=1\;$}
\def\ntwot{${\mathcal N}=(2,2)\;$}

\def\ba{\beq\new\begin{array}{c}}
\def\ea{\end{array}\eeq}
\def\be{\ba}
\def\ee{\ea}

\begin{document}

\title[]{Non-Abelian Strings: From Weak to Strong Coupling and Back via Duality}
\author{M.~Shifman and A. Yung}{address={William I. Fine Theoretical Physics Institute,
University of Minnesota,
Minneapolis, MN 55455, USA\\Petersburg Nuclear Physics Institute, Gatchina, St. Petersburg
188300, Russia}}

\keywords{Nonperturbative supersymmetry, duality}

\classification{11.30.Pb, 11.27.+d}

\begin{abstract}

The crossover transition
from  weak coupling  at large $\xi$ to  strong coupling  at small $\xi$
is studied in
 \ntwo supersymmetric gauge theory with the U($N$) gauge group
and  $N_f>N $  (here $\xi$ is the Fayet--Iliopoulos parameter).
We find that at strong coupling a dual non-Abelian weakly coupled ${\mathcal N}=2$
theory exists which describes low-energy physics at small $\xi$.
The dual gauge group is U$(N_f-N)$. The dual theory has $N_f$ flavors
of light dyons, to be compared with $N_f$ quarks in the original U($N$)
theory. 
Both theories support non-Abelian semilocal strings.
 In each of these two regimes there are two
  varieties of physical excitations: elementary fields and nonperturbative composite states bound by confining strings.
These varieties interchange upon transition from one regime to the other. 
We conjecture that the composite stringy states can be related to Seiberg's $M$ fields.

\end{abstract}
\maketitle

Duality is one the most powerful methods in modern theoretical  physics. It allows one to study a theory at strong coupling using its dual formulation which at weak coupling.
In this talk new results  \cite{SYdual} on non-Abelian duality will be reported.

First, let us ask ourselves what we learned about duality in four-dimensi\-onal gauge
theories in the last 15 years. 
The summary is as follows: 

(i) 
the Seiberg--Witten electromagnetic duality in \ntwo SQCD  \cite{SW1,SW2} 
 reduces, in the infrared,  the underlying non-Abelian
theory  to an Abelian effective theory.  Condensation of monopoles 
(charges in the dual theory)
leads to formation of 
the Abrikosov--Nielsen--Olesen (ANO) flux tubes (strings).
The flux they carry is that of the magnetic field of the dual theory,
which is equivalent to the electric flux in the original theory. Thus,
they confine quarks.  Confinement is essentially Abelian.

(ii)
The second example is   Seiberg's duality in \none
SQCD \cite{Sdual}. It connects the underlying SU($N$) theory with $N_f$ flavors with
its dual counterpart,
having the SU($\tN$) gauge group
 ($\tN=N_f-N$) and  $N_f$ flavors of ``dual quarks"
coupled to a neutral mesonic field $M$. Seiberg's duality also takes place in the infrared;
the ultraviolet behavior of the dual partners is different.

The common belief is that  as we deform \ntwo SQCD adding a mass term for the
adjoint matter, $\mu {\rm Tr} \Phi^2$, the Seiberg--Witten Abelian duality smoothly goes into Seiberg's non-Abelian duality.
The monopoles of (i) become ``dual quarks" of (ii),  and  their condensation leads to non-Abelian confinement of quarks.

\begin{figure}
        \includegraphics[width=0.3\textwidth,height=1.2in]{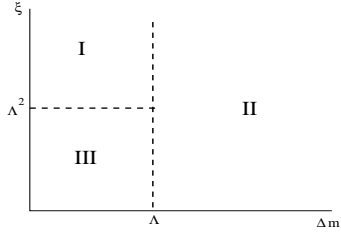}
        \caption{Various regimes in  \ntwo QCD are  separated by crossovers.}
        \label{figphasediag}
\end{figure}

The results we report \cite{SYdual} (see also \cite{SYcross}) extend those of \cite{SYdual}; they 
do {\em not} fully support the above common belief. We find:
\begin{itemize}
\item Duality between two non-Abelian theories.
Non-Abelian duality is not the electromagnetic duality.
Monopoles are confined in both original and dual theories;
 \item Both theories in the dual pair have non-Abelian strings which confine;
 \item We observe small$\leftrightarrow$large $\xi$ duality in the bulk and on 
    the string world sheet;
  \item Weakly coupled domain is separated from 
 strongly coupled by a crossover transition with respect to $\xi$.
\end{itemize}
Here $\xi$ is the Fayet--Iliopoulos (FI) term,
a key element of our set-up which can be summarized as follows.
\begin{figure}
        \includegraphics[width=0.25\textwidth,height=0.6in]{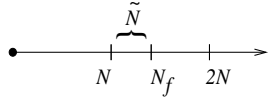}
        \caption{The number of flavors exceeds the number of colors,
$\tilde N\equiv N_f - N >0.$}
        \label{voenn}
\end{figure}
We consider  \ntwo SQCD with the  gauge group U$(N)$ assuming  
$N_f>N$ but   $N_f<2N$ to keep asymptotic freedom in the original theory, see Fig. \ref{voenn}.
The Fayet--Iliopoulos  term $\xi\neq 0$
triggers condensation of $N$ squark fields. 
The (s)quark mass differences 
\beq
\Delta m_{AB}=m_A-m_B\,,\qquad A,B=1,...,N_f
\label{sqmd}
\eeq
are variable parameters.
If $\xi$ is large,  the theory is at weak coupling,
while at small $\xi$ it becomes strongly coupled.
A weakly coupled dual   description is constructed 
at small $\xi$.
Various regimes  of the theory in the $\{ \xi,\,\Delta m\}$ plane are schematically shown 
in Fig.~\ref{figphasediag}. 

The bosonic part of our basic 
theory has the form  
\beqn
S&=&\int d^4x \left[\frac1{4g^2_2}
\left(F^{a}_{\mu\nu}\right)^2 +
\frac1{4g^2_1}\left(F_{\mu\nu}\right)^2
+
\frac1{g^2_2}\left|D_{\mu}a^a\right|^2 +\frac1{g^2_1}
\left|\partial_{\mu}a\right|^2 \right.
\nonumber\\[4mm]
&+&\left. \left|\nabla_{\mu}
q^{A}\right|^2 + \left|\nabla_{\mu} \bar{\tilde{q}}^{A}\right|^2
+V(q^A,\tilde{q}_A,a^a,a)\right],
\label{model}
\eeqn
where
\beqn
&& 
V(q^A,\tilde{q}_A,a^a,a) =
 \frac{g^2_2}{2}
\left( \frac{1}{g^2_2}\,  f^{abc} \bar a^b a^c
 +
 \bar{q}_A\,T^a q^A -
\tilde{q}_A T^a\,\bar{\tilde{q}}^A\right)^2
\nonumber\\[3mm]
&&
+ \frac{g^2_1}{8}
\left(\bar{q}_A q^A - \tilde{q}_A \bar{\tilde{q}}^A -N \xi\right)^2
+ 2g^2_2\left| \tilde{q}_A T^a q^A \right|^2+
\frac{g^2_1}{2}\left| \tilde{q}_A q^A  \right|^2
\nonumber\\[3mm]
&&
+\frac12\sum_{A=1}^{N_f} \left\{ \left|(a+\sqrt{2}m_A +2T^a a^a)q^A
\right|^2 
+ 
\left|(a+\sqrt{2}m_A +2T^a a^a)\bar{\tilde{q}}^A
\right|^2 \right\}.
\nonumber
\eeqn

With
degenerate quark masses $\Delta m_{AB}=0$,
the microscopic  theory  has an unbroken global SU$(N)$ symmetry
which is a diagonal combination of SU$(N)_{\rm color}$
and an SU$(N)$ subgroup of the flavor SU$(N_f)$ group acting in the theory.
Thus, the color-flavor locking takes place.  All light states come in the adjoint and singlet 
representations of the unbroken SU$(N)_{C+F}$. 

\begin{figure}
        \includegraphics[width=0.5\textwidth,height=1.3in]{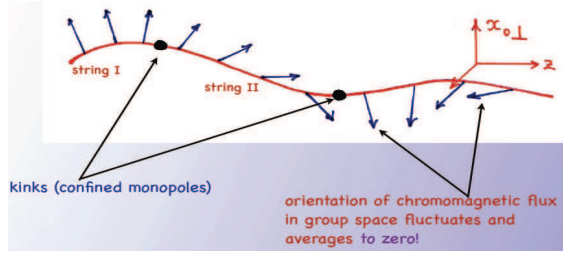}
        \caption{Non-Abelian strings.}
        \label{three}
\end{figure}

The  theory  supports
non-Abelian strings \cite{HT1,SYmon}
(for reviews see \cite{SYrev}). Since  $N_f >N $,  these strings are semilocal
and would not lead to linear confinement
if all $\Delta m$'s were set to zero. A pictorial representation of non-Abelian strings
is given in Fig.~\ref{three}.
 
In the original theory the squark fields are condensed in domain I, Fig.~\ref{figphasediag}. The theory is fully Higgsed. The
 monopoles are attached to strings. In fact, the SU$(N)$ monopoles  represent 
 the junctions between two distinct degenerate strings. They
  are seen as kinks in the world-sheet sigma model \cite{SYmon},
  see Fig.~\ref{three}.  

The domain II is that of the {\em Abelian} Higgs regime at weak coupling. As we increase  
  $\Delta m_{AB}$, the (off diagonal) $W$ bosons and their superpartners become exceedingly 
heavier and decouple from the low-energy spectrum. We are  left with the
photon (diagonal) gauge fields and their quark \ntwo superpartners. Explicit
breaking of the flavor symmetry by $\Delta m_{AB}\neq 0$ leads to the loss
of the non-Abelian nature of the string solutions; they  become
Abelian (the so-called $Z_N$) strings. 

Finally, as we reduce $\xi$ and  $|\Delta m_{AB}|$ below $\Lambda$, we enter the strong coupling 
domain III. We show that at $N_f>N$ there is a dual description
in this domain, moreover, the dual theory is non-Abelian, with 
 the dual gauge group
\beq
{\rm U}(\tN)\times {\rm U}(1)^{N-\tN}\,,
\label{dualgaugegroup}
\eeq
and $N_f$ flavors of  charged non-Abelian dyons.  In its gross features the dual \ntwo theory
that we found is
similar to Seiberg's dual \cite{Sdual}  to our original microscopic theory. 
Because
$N_f>2\tN$,  the dual theory is infrared (IR) free rather than
asymptotically free. This result is in perfect match with the
results obtained in \cite{APS} where the
dual non-Abelian gauge group SU$(\tN)$ was identified at the root of a
baryonic branch in the SU$(N)$ gauge theory with massless quarks (see also \cite{CKM}).

In  the limit of 
degenerate quark masses $\Delta m_{AB}=0$ and small $\xi$,
the dual theory has an unbroken global diagonal SU$(\tN)$ symmetry.
It is obtained as a result of the spontaneous breaking of the gauge U$(\tN )$
group and an SU$(\tN )$ subgroup of the flavor SU$(N_f)$ group. 
Thus, the color-flavor locking takes place in the dual theory as well, much in the same way
as in the original microscopic theory in the domain I, albeit the preserved diagonal symmetry is different.
The light states come in adjoint and singlet representations of the global SU$(\tN )_{C+F}$. Thus, the low-energy spectrum of the theory
in the domain III is dramatically different from that of domain I. Excitation spectra are arranged in
different representations of the global unbroken groups, SU$(N)$ and SU$(\tN )$, respectively.

Three above-mentioned regimes  --- three domains shown in Fig.~\ref{figphasediag} ---
are arguably separated by crossovers, much in the same way
as it happens in the case $N_f=N$ \cite{SYcross}. 
 The evidence in favor of  crossovers (rather than phase transitions)  can 
 be summarized as follows.
 
\begin{itemize}

\item  In the equal quark mass limit the domains I and III have Higgs branches of the same dimensions
and the same pattern of global symmetry breaking;

\item
For generic masses $\Delta m_{AB}\neq 0$  all three regimes have the same number of isolated vacua 
at nonvanishing $\xi$;

\item Each of these vacua has the same number ($=N$) of distinct elementary strings in all three
domains. Moreover, the
BPS spectra of excitations on the non-Abelian string coincide in the domains I and III.

\end{itemize}

\section{Basic features of the set-up}

The field content is as follows. The \ntwo vector multiplet
consists of the  U(1)
gauge field $A_{\mu}$ and the SU$(N)$  gauge field $A^a_{\mu}$,
where $a=1,..., N^2-1$, and their Weyl fermion superpartners
 plus
complex scalar fields $a$, and $a^a$ and their Weyl superpartners.
The $N_f$ quark multiplets of  the U$(N)$ theory consist
of   the complex scalar fields
$q^{kA}$ and $\tilde{q}_{Ak}$ (squarks) and
their   fermion superpartners, all in the fundamental representation of 
the SU$(N)$ gauge group.
Here $k=1,..., N$ is the color index
while $A$ is the flavor index, $A=1,..., N_f$. We will treat $q^{kA}$ and $\tilde{q}_{Ak}$
as rectangular matrices with $N$ rows and $N_f$ columns.

Let us discuss  the vacuum structure of  this theory.
The  vacua of the theory (\ref{model}) are determined by the zeros of 
the potential $V$. With the generic choice of the quark masses we have 
$C_{N_f}^{N}= N_f!/N!\tN!$ isolated $r$-vacua in which $r=N$ quarks (out of $N_f$) develop
vacuum expectation values  (VEVs).
Consider, say, the (1,2,...,$N$) vacuum in which the first $N$ flavors develop VEVs.
We can exploit gauge rotations to
make  all squark VEVs real. Then
in the problem at hand they take the form
\beq
\langle q^{kA}\rangle =\sqrt{
\xi}\,
\left(
\begin{array}{cccccc}
1 & \ldots & 0 & 0 & \ldots & 0\\
\ldots & \ldots & \ldots  & \ldots & \ldots & \ldots\\
0 & \ldots & 1 & 0 & \ldots & 0\\
\end{array}
\right),
\qquad \langle \bar{\tilde{q}}^{kA}\rangle =0,
\label{qvev}
\eeq
where we write down the quark fields as  matrices in color and flavor indices
($k=1,..., N\,,\,\, A=1,...,N_f$).
The FI term $\xi$ singles  out the $r=N$ vacua from the set of all $r$-vacua. 

In the vacuum under consideration the
adjoint fields also develop  
VEVs, namely,
\beq
\left\langle \left(\frac12\, a + T^a\, a^a\right)\right\rangle = - \frac1{\sqrt{2}}
 \left(
\begin{array}{ccc}
m_1 & \ldots & 0 \\
\ldots & \ldots & \ldots\\
0 & \ldots & m_N\\
\end{array}
\right),
\label{avev}
\eeq
For generic values of the quark masses, the  SU$(N)$ subgroup of the gauge 
group is
broken down to U(1)$^{N-1}$. However, in the special limit
\beq
m_1=m_2=...=m_{N_f},
\label{equalmasses}
\eeq
the  SU$(N)\times$U(1) gauge group remains  unbroken by the adjoint field.
In this limit the theory acquires a global flavor SU$(N_f)$ symmetry.

While the adjoint VEVs do not break the SU$(N)\times$U(1) gauge group in the limit
(\ref{equalmasses}), the quark condensate (\ref{qvev}) results in  the spontaneous
breaking of both gauge and flavor symmetries.
A diagonal global SU$(N)$ combining the gauge SU$(N)$ and an
SU$(N)$ subgroup of the flavor SU$(N_f)$
group survives, however. We refer to this diagonal
global symmetry as to $ {\rm SU}(N)_{C+F}$.
The color-flavor locking takes place  in a slightly different way 
than in the case $N_f = N$ (or $\tilde N =0$).
The presence of the global SU$(N)_{C+F}$ group is instrumental for
formation of the non-Abelian strings.

More exactly, the pattern of breaking of the
color and flavor symmetry 
is as follows: 
\beq
{\rm U}(N)_{\rm gauge}\times {\rm SU}(N_f)_{\rm flavor}\to  {\rm SU}(N)_{C+F}\times  {\rm SU}(\tilde{N})_F\times {\rm U}(1)\,,
\label{c+f}
\eeq
where $\tilde{N}=N_f-N$.
Here  SU$(\tN )_F$ factor stands for the flavor rotation of the 
$\tN$ quarks. For unequal quark masses the  global symmetry  (\ref{c+f}) is broken down to 
U(1)$^{N_f-1}$.

All
gauge bosons in the bulk are massive,
\beq
m_{\gamma} =g_1\,\sqrt{\frac{N}{2}\,\xi\,},\qquad m_{W}=g_2\sqrt{\xi}.
\label{phmass}
\eeq
The adjoint fields $a$ and
$a^a$ as well as $N^2$ components of the quark matrix $q$ acquire
the same masses as the corresponding gauge bosons. 

\section{A journey through various regimes}

\begin{figure}
        \includegraphics[width=0.65\textwidth,height=1.1in]{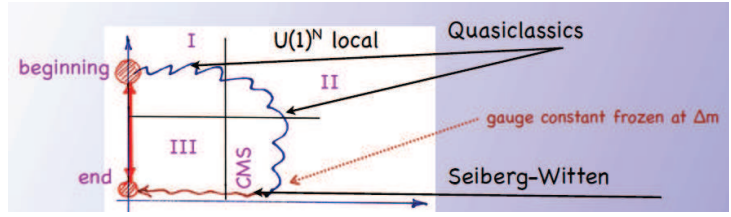}
        \caption{The transitions  in the $\{\xi ,\,\, \Delta M_{\rm inside}\}$ plane.}
        \label{four}
\end{figure}

We will single out a group of $N$ quarks and another one of $\tN$ quarks.
We generically refer to the masses in the first and second groups as
$m_P$  and $m_K$, respectively;
$P=1, ..., N$ numerates the quark flavors which develop
VEVs, while $K=N+1, ..., N_f$ numerates ``extra" quark flavors. 
The extra flavors become massless in the
limit (\ref{equalmasses}).
The mass differences inside the first group
(or inside the second group) are called $\Delta M_{\rm inside}$.
The mass differences $m_P-m_K$ are referred to as 
$\Delta M_{\rm outside}$.  The transitions we will study are those in $\xi$ (the vertical axis)
and $\Delta M_{\rm inside}$ (the horizontal axis in 
Fig.~\ref{four}).
At $\xi =0$ we arrive at the Seiberg--Witten solution. Note that
$m_P-m_K\equiv \Delta M_{\rm outside}$ is kept fixed in the process. 
Eventually, we take $\Delta M_{\rm outside}\ll\Lambda$. 
This is necessary to get the dual theory non-Abelian.

If $m_P-m_K\equiv \Delta M_{\rm outside}\neq 0$, the extra quark
flavors acquire masses determined by the mass differences
$m_P-m_K$.

Note that all states come in representations of the unbroken global
 group (\ref{c+f}), namely, the singlet and adjoint representations
of SU$(N)_{C+F}$
\beq
(1,\, 1), \quad (N^2-1,\, 1),
\label{onep}
\eeq
 and bifundamentals
\beq
 \quad (\bar{N},\, \tN), \quad
(N,\, \bar{\tN})\,,
\label{twop}
\eeq
 where we mark representation with respect to two 
non-Abelian factors in (\ref{c+f}).

\begin{figure}
        \includegraphics[width=0.25\textwidth,height=0.5in]{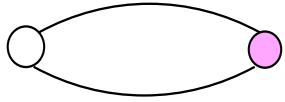}
        \caption{Meson formed by antimonopole and dyon connected by two strings.
Open and closed circles denote dyon and antimonopole, respectively.
}
        \label{figmeson}
\end{figure}

In the beginning we have the gauge group U$(N)$, with $N_f$ matter hypermultiplets.
The light part of the spectrum includes the vector supermultiplet 
($16\times (N^2-1)$ degrees of freedom with mass $\sim g\sqrt \xi$)
plus extra bifundamentals (\ref{twop}). In addition, there are $\tN^2-1$ composites
of the type presented in Fig.~\ref{figmeson}.
Their mass is heavy (i.e. $\sim \sqrt\xi$). In the end we 
have the gauge group U$(\tN)$, with $\tN$ matter hypermultiplets.
The light part of the spectrum includes the vector supermultiplet 
($16\times (\tN^2-1)$ degrees of freedom with mass $\sim g\sqrt \xi$).
In addition, there are extra bifundamentals (\ref{twop}) and $N^2-1$ heavy composites $D\bar D$.

\section {Non-Abelian strings}
\label{strings}

The  $Z_N$-string solutions in the theory with $N_f=N$
break the  SU$(N)_{C+F}$ global group. Therefore,
strings have orientational zero modes, associated with rotations of their color
flux inside the non-Abelian SU($N$). 
The global group is broken  down to
${\rm SU}(N-1)\times {\rm U}(1)$.
As a result,
the moduli space of the non-Abelian string is described by the coset space
\beq
\frac{{\rm SU}(N)}{{\rm SU}(N-1)\times {\rm U}(1)}\sim CP(N-1)\,.
\label{modulispace}
\eeq
The low-energy effective theory on the world sheet of the non-Abelian string
is \ntwo SUSY two-dimensional $CP(N-1)$ model \cite{HT1,SYmon}.

Now   we add $\tN$ ``extra" quark flavors (first, with degenerate masses).
Then the strings  become semilocal, whose transverse size is a modulus.
In this case these strings do not generate linear confinement.
However, at the end, $\Delta M_{\rm outside}\neq 0$
will lift the size moduli, so that linear confinement will ensue.

Non-Abelian semilocal strings  have two types of moduli: orientational and size moduli.
The orientational
zero modes  are parametrized by a complex vector $n^P$, $P=1,...,N$,
 while its $\tN=(N_f-N)$ size moduli are parametrized by another complex vector
$\rho^K$, $K=N+1,...,N_f$. The effective two-dimensional theory
which describes  internal dynamics of the non-Abelian semilocal string is
an \ntwot ``toric" sigma model   including both types of fields. Its bosonic action
in the gauge formulation (which assumes taking the limit $e^2\to\infty$)
has the form
\beqn
&&S = \int d^2 x \left\{
 \left|\nabla_{\alpha} n^{P}\right|^2 
 +\left|\tilde{\nabla}_{\alpha} \rho^K\right|^2
 +\frac1{4e^2}F^2_{\alpha\beta} + \frac1{e^2}\,
\left|\pt_{\alpha}\sigma\right|^2
\right.
\nonumber\\[3mm]
&+&\left.
2\left|\sigma+\frac{m_P}{\sqrt{2}}\right|^2 \left|n^{P}\right|^2 
+ 2\left|\sigma+\frac{m_{K}}{\sqrt{2}}\right|^2\left|\rho^K\right|^2
+ \frac{e^2}{2} \left(|n^{P}|^2-|\rho^K|^2 -2\beta\right)^2
\right\},
\nonumber\\[4mm]
&& 
P=1,...,N\,,\qquad K=N+1,...,N_f\,,\qquad \tilde{\nabla}_k=\pt_k+iA_k\,.
\label{wcp}
\eeqn
The fields $n^{P}$ and $\rho^K$ have
charges  +1 and $-1$ with respect to the auxiliary U(1) gauge field,
hence, the difference in the covariant derivatives, $ \nabla_i=\d_i-iA_i$
and $\tilde{\nabla}_j=\d_j+iA_j$ respectively.

The $D$-term condition
\beq
  |n^P|^2 - |\rho^K|^2=2\beta\,,
\label{unitvec}
\eeq
is implemented in the limit $e^2\to\infty$. Moreover, in this limit
the gauge field $A_{\alpha}$  and its \ntwo bosonic superpartner $\sigma$ become
auxiliary and can be eliminated.
The  two-dimensional coupling constant $\beta$ is related to the four-dimensional
one as
$
\beta= {2\pi}/{g_2^2}\,.
$

\section{Monodromies, or what becomes of the quark fields in the journey}
\label{bulkdual}

To simplify presentation, we will consider a particular example, $N=3$ and $\tN=2$.
In analyzing the transition  from   domain I to 
III (see Fig.~\ref{figphasediag}) we make  two steps.
First,
we take  the quark mass differences to be large, passing to   domain II. In this domain the theory stays at
weak coupling, and we can safely decrease the value of the FI parameter $\xi$. Next,
we   use the exact Seiberg--Witten solution of the theory on the Coulomb branch \cite{SW1,SW2}
(i.e. at   $\xi=0$) to perform the passage from  domain II to  III. 

\begin{figure}
        \includegraphics[width=0.55\textwidth,height=0.7in]{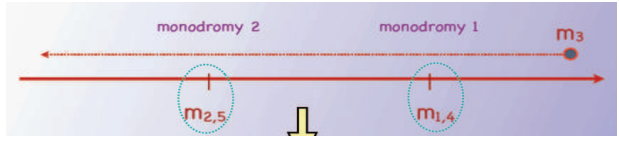}
        \caption{Two Argyres--Douglas points on the way from domain I to III.
}
        \label{five}
\end{figure}

In this journey we will have to carefully consider two Argyres--Douglas points, to deal with two monodromies,
as we vary $m_3$, see Fig.~\ref{five}.
In each passage we split the masses, first $m_1$ and $m_4$, and then $m_2$ and $m_5$.
At the end we tend $m_1\to m_2\to m_3$ and $m_5\to m_4$

We  investigate the monodromies
using the  approach of Ref.~\cite{SYcross} which is  similar to that of Ref.~\cite{CKM}. We start with 
the $r=3$ vacuum at large 
$\xi$ in domain I, where three quarks with charges
\beqn
&&\left(n_e,n_m;\,n_e^3,n_m^3;\,n_e^8,n_m^8\right)=
\left(\frac12,0;\,\frac12,0;\,\frac1{2\sqrt{3}},0\right), \nonumber\\[2mm]
&&\left(n_e,n_m;\,n_e^3,n_m^3;\,n_e^8,n_m^8\right)=
\left(\frac12,0;\,-\frac12,0;\,\frac1{2\sqrt{3}},0\right), \nonumber \\[2mm]
&&\left(n_e,n_m;\,n_e^3,n_m^3;\,n_e^8,n_m^8\right)=
\left(\frac12,0;\,0,0;\,-\frac1{\sqrt{3}},0\right),
\label{quarkcharges}
\eeqn
develop VEV's. Here
$n_e$ and $n_m$ denote electric and 
magnetic charges of a given state with respect to the U(1) gauge group, while $n_e^3$,
$n_m^3$ and $n_e^8$, $n_e^8$ stand for the electric and magnetic charges  with respect to the Cartan
generators of the SU(3) gauge group (broken down to U(1)$\times$U(1) by $\Delta m_{AB}$).

Then in domain III these quarks transform into light
dyons with charges
\beqn
&& D^{11}:\,\,\, \left(\frac12,0;\,\frac12,\frac12;\,\frac1{2\sqrt{3}},\frac{\sqrt{3}}{2}\right),
\nonumber\\[2mm]
&& D^{22}:\,\,\, 
\left(\frac12,0;\,-\frac12,-\frac12;\,\frac1{2\sqrt{3}},\frac{\sqrt{3}}{2}\right),
\nonumber\\[2mm]
&& D_{33}:\,\,\, \,
\left(\frac12,0;\,0,0;\,-\frac1{\sqrt{3}},-\sqrt{3}\right).
\label{dyons}
\eeqn

 For  consistency of
our analysis it is instructive to
consider another route from the domain I to the domain III, namely the one along the line
 $\Delta M_{\rm inside}=0$. On this 
line we keep the global color-flavor locked group  unbroken. Then we obtain a surprising result:
the quarks and gauge bosons
which form the  adjoint $(N^2-1)$ representation  of SU($N$) at large $\xi$ and the dyons and gauge bosons which form the  adjoint $(\tN^2-1)$ representation  of SU($\tN$) at small $\xi$ are, in fact, {\em distinct} states.
How can this occur?

Since we have a crossover  between the domains I and III rather than  a phase 
transition,
this means that in the full microscopic theory the $(N^2-1)$  adjoints of SU($N$) become heavy 
and decouple as we pass from the
domain I to III along the line $\Delta m_{AB}=0$. Moreover, some 
composite $(\tN^2-1)$ adjoints  of SU($\tN$), which are 
heavy  and invisible in the low-energy description in the domain I become light in 
the  domain III and form the $D^{lK}$ dyons
 ($K=N+1,...,N_f$) and gauge bosons $B^p_{\mu}$. The phenomenon of level crossing
 takes place. Although this crossover is smooth in the full theory,
from the standpoint of the low-energy description the passage from  the domain I to 
 III means a dramatic change: the low-energy theories in these domains are 
completely
different; in particular, the degrees of freedom in these theories are different.

This logic leads us to the following conclusion. In addition to light dyons and gauge bosons 
 included in  the low-energy theory  in the domain III at small $\xi$, we have
heavy  fields (with masses of the order of $\Lambda$) which form the adjoint representation
$(N^2-1,1)$ of the global symmetry. These are screened (former)  quarks 
and gauge bosons from the domain I continued into III.
 Let us denote them as $M_P^{P'}$ ($P,P'=1,...,N$). What is their physical nature in the region III?
 
\begin{figure}
        \includegraphics[width=0.27\textwidth,height=1in]{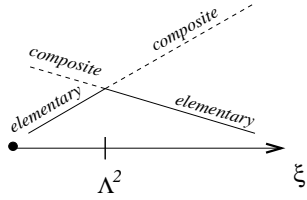}
        \caption{Evolution of the SU$(N)$ and SU$(\tN)$ $W$ bosons vs. $\xi$. 
}
        \label{figevol}
\end{figure}

Before answering this question let us note that
by the same token, it is  seen that in domain I, in addition to the light quarks and gauge bosons,
we  have heavy fields $M_K^{K'}$ ($K,K'=N+1,...,N_f$), which  form the  adjoint $(\tN^2-1)$ representation  of SU($\tN$).
This is schematically depicted in Fig.~\ref{figevol}.

Now we come back to the physical nature of adjoint fields $M_P^{P'}$ in the region III.
It is well known that the $W$
bosons usually do not exist as localized states 
 in the strong coupling regime on the Coulomb branch (speaking in jargon, they ``decay"). They split
into antimonopoles and dyons on CMS on  which   the Argyres--Douglas  points lie \cite{SW1,BF}.

 Consider, for example, the $W$-boson associated with the
$T^3$  generator ($T^3$ $W$ boson for short) with the charge
$(0,0;\,1,0;\,0,0)$ in the domain II.
As we go through CMS this $W$ boson  decay
into the $T^3$ antimonopole and 
dyon with the charges $(0,0;\, 0,-1;\,0,0)$ and $(0,0;\,1,1;\,0,0)$, respectively. 
It means that the $W$ boson is absent 
 in domain III,  in full accord with the analysis of
the SU(2) theory in \cite{BF}.

This picture is valid on the Coulomb branch at $\xi=0$.  As we switch on small $\xi\neq 0$
the  monopoles
and dyons become confined by strings. In fact, the elementary monopoles/dyons are represented by junctions
of two different elementary non-Abelian strings \cite{T,SYmon}, see also a detailed 
discussion of the monopole/dyon confinement in \cite{SYdual}. This
means that, as we move from the domain II into  III at small nonvanishing $\xi$ the
$W$ boson ``decays" into an antimonopole and dyon; however,  these states cannot 
abandon each other and move far apart because they
are confined. Therefore, the
$W$ boson evolves into a stringy meson formed by an antimonopole and dyon connected by
two strings, as shown in  Fig.~\ref{figmeson}, see \cite{SYrev} for a discussion of these
stringy mesons.

These stringy mesons
have nonvanishing U(1) global charges with respect to the Cartan generators of the SU(3) subgroup
of the global group (\ref{c+f}) (above we discussed  only one $W$ boson of this type, related to the
$T^3$ generator, however, in fact, we have six different charged gauge boson/quark states of this type). 
In the equal
mass limit these globally charged stringy mesons combine with neutral (with respect to the group
U(1)$^{N_f-1}$ stringy mesons formed by pairs of monopoles and antimonopoles (or
dyons and antidyons) connected by two strings, to
form the octet representation  of the SU(3) subgroup
of the global group (\ref{c+f}) (in general, the adjoint representation of SU$(N)$). 
They are heavy  in the domain III, with mass of the order of $\Lambda$.

We  identify these stringy mesons with $(N^2-1)$ adjoints 
 $M_P^{P'}$ ($P,P'=1,...,N$)
 of the SU$(N)$ subgroup with which we have seen
{\em en route} from the domain I to  III along the line $\Delta m_{AB}=0$.

The same applies to the $q^{kK}$ quarks  ($K=N+1,...,N_f$)
of the domains I and II. As we go through the crossover
into the domain III at small $\xi$ $q^{kK}$ quarks evolve into stringy mesons formed by pairs of antimonopoles
and dyons connected by two strings, see Fig.~\ref{figmeson}. However, these states are unstable.
To see that this is indeed the case, please, observe
 that in the equal mass limit
these stringy mesons  fill the bifundamental representations $(N,\bar{\tN})$ and $(\bar{N},\tN)$ 
of the global group; hence, can decay into light dyons/dual gauge bosons with the same
quantum numbers.

It is quite plausible to suggest that these fields $M_P^{P'}$ and $M_K^{K'}$ are Seiberg's mesonic fields
\cite{Sdual,IS},
which occur in the dual theory upon breaking of \ntwo supersymmetry by the mass-term
 superpotential $\mu{\mathcal A}^2$ for the adjoint fields when we take the limit 
$\mu\to\infty$. In this limit  our theory becomes \none SQCD. Previously, these $M_{AB}$ fields were not identified 
in the \ntwo theory.

In conclusion, we demonstrated that non-Abelian confinement in our theory  is a 
combined effect of the Higgs screening, ``decay" processes on CMS and confining string
formation. The 
strings that are dynamically formed always 
confine monopoles or dyons (whose charges can be represented as a sum of those of a monopole 
plus  $W$-boson)
both, in  the original and dual theories, rather than quarks.

\end{document}